\documentclass[12pt,leqno]{article}   
\usepackage[page,toc,titletoc,title]{appendix}
\usepackage{tocloft}
\usepackage{amsmath,amssymb}
\usepackage{blindtext}
\usepackage{lipsum}
\usepackage{tabularx}
\usepackage{booktabs}
\usepackage{rotating}
\usepackage{array}
\usepackage{pdflscape}
\usepackage{caption}
\usepackage{tabularray}
\UseTblrLibrary{booktabs}
\usepackage{blochsphere}
\usepackage{caption}  
\usepackage[T1]{fontenc}
\usepackage[utf8]{inputenc}
\usepackage{authblk}
\usepackage{lmodern}
\usepackage{mathtools}
\usepackage[all,cmtip]{xy}
\usepackage[all]{xy}
\usepackage{tikz-cd}
\usepackage{braket}
\usepackage{enumerate}
\usepackage{wrapfig}
\usepackage{quantikz}
\usepackage{pgf,tikz,pgfplots}
\usepackage{amsmath,amstext,amsthm,amsfonts,amssymb,color,latexsym}
\makeatletter 
\tikzset{ 
	reuse path/.code={\pgfsyssoftpath@setcurrentpath{#1}} 
} 
\tikzset{even odd clip/.code={\pgfseteorule}, 
	protect/.code={ 
		\clip[overlay,even odd clip,reuse path=#1] 
		(current bounding box.south west) rectangle (current bounding box.north east)
		; 
}} 
\makeatother 
\usetikzlibrary{3d,perspective,angles,quotes}
\pgfmathsetmacro{\myaz}{10}
\numberwithin{equation}{section}

\makeatletter
\newcommand\appendix@section[1]{%
	\refstepcounter{section}%
	\orig@section*{Appendix \@Alph\c@section: #1}%
	\addcontentsline{toc}{section}{Appendix \@Alph\c@section: #1}%
}
\let\orig@section\section
\g@addto@macro\appendix{\let\section\appendix@section}
\makeatother

\usepackage{epsfig}

\newcommand{\beano}{\begin{eqnarray*}}
	\newcommand{\enano}{\end{eqnarray*}}
\newcommand{\ena}{\end{eqnarray}}

\newcommand{\be}{\begin{equation}}
\newcommand{\ee}{\end{equation}}
\newcommand{\en}{\end{equation}}

\newcommand{\ba}{\begin{array}}
\newcommand{\ea}{\end{array}}

\newcommand{\bg}{\begin{gathered}}
\newcommand{\eg}{\end{gathered}}

\newcommand{\bea}{\begin{eqnarray}}
\newcommand{\eea}{\end{eqnarray}}

\title{Qubit Geometry through Holomorphic Quantization}
\author[1]{Ahmad Hazazi Ahmad Sumadi \thanks{Corresponding author:
ahmadhazazi@gmail.com}}
\author[1,2]{Nurisya Mohd Shah \thanks{Corresponding author: risya@upm.edu.my}}
\author[1,3]{Umair Abdul Halim} 
\author[4]{Hishamuddin Zainuddin}

\affil[1]{Institute for Mathematical Research (INSPEM), Universiti Putra Malaysia (UPM), 43400 UPM Serdang, Selangor Darul Ehsan,  MALAYSIA}
\affil[2]{Department of Physics, Universiti Putra Malaysia (UPM), 43400 UPM Serdang, Selangor Darul Ehsan,  MALAYSIA}
\affil[3]{Center for Foundation Studies in Science of Universiti Putra Malaysia (ASPutra), Universiti Putra Malaysia (UPM), 43400 UPM Serdang, Selangor Darul Ehsan,  MALAYSIA}
\affil[4]{School of Mathematics and Physics, Xiamen University Malaysia, Jalan Sunsuria, Bandar Sunsuria, 43900, Sepang, Selangor,  MALAYSIA }

\begin{document}

\maketitle
\begin{abstract}
We develop a wave mechanics formalism for qubit geometry using holomorphic functions and M\"obius transformations, providing a geometric perspective on quantum computation. This framework extends the standard Hilbert space description, offering a natural interpretation of standard quantum gates on the Riemann sphere that is examined through their M\"obius action on holomorphic wavefunction. These wavefunctions emerge via a quantization process, with the Riemann sphere serving as the classical phase space of qubit geometry. We quantize this space using canonical group quantization with holomorphic polarization, yielding holomorphic wavefunctions and spin angular momentum operators that recover the standard $SU(2)$ algebra with interesting geometric properties. Such properties reveal how geometric transformations induce quantum logic gates on the Riemann sphere, providing a novel perspective in quantum information processing. This result provides a new direction for exploring quantum computation through Isham's canonical group quantization and its holomorphic polarization method.

\vspace{.3in}


{\bf Keywords and phrases:} Canonical group quantization, compact phase space, holomorphic polarization, qubit geometry, qubit operations.
\end{abstract}
\section{Introduction }


Quantum information processing (QIP) represents one of the most promising frontiers in information technology, offering transformative potential in computation \cite{Shor, Harrow}, secure communication \cite{Wiesner}, cryptography \cite{Bennett}, and sensing and metrology \cite{Lloyd}. At its core, QIP involves the manipulation of quantum states, typically represented as vectors or density matrices in a Hilbert space \cite{Chuang}. However, beyond their algebraic descriptions, these states inhabit rich geometric structures that profoundly influence their dynamics and applications. Geometry provides a natural framework for understanding the properties of quantum states and operations \cite{Kibble}. The state spaces of quantum systems are not merely mathematical abstractions; they possess symplectic structures, Riemannian metrics, and topological features that dictate the evolution and interaction of quantum states \cite{Brody, Aharanov, Schilling}. Tools from geometry facilitate insights into entanglement, quantum error correction, and computational algorithms, offering powerful approaches to problems in quantum information science \cite{Fioresi}. \newline

The basic unit of information in quantum computers is based on a two-level quantum system called a qubit, which serve the same function as a classical bit in a classical computer. In this study, we will investigate the idea of wave mechanics formalism of qubit geometry, for which the latter is not new \cite{Mosseri1,Mosseri2, Bengtsson, Bengsston2,  Hiley, Kus, Levay, Isidro}. The main idea of our study is to reveal the formalism of qubit in the realm of traditional wave mechanics formalism of quantum theory through the process of quantization, which is a well-known and very useful technique in theoretical physics, for instance in quantum gravity \cite{Renate} and quantum field theory \cite{Witten}. In particular, the idea is that, qubits and qubit operations are often represented elementarily as matrices and there seems to be a disconnect between such formalism with the traditional wave mechanics formalism of quantum theory.  We reconnect with the wave formalism through the process of quantization and there are advantages of having the various different representations \cite{Chabaud}. The starting point is to recognize that the projective Hilbert space as a classical phase space \cite{Bjelakov} For finite-dimensional Hilbert spaces, these will be complex projective spaces and in particular, for a single qubit is the Riemann sphere (or Bloch sphere) \cite{Urbantke, Lee}.  It is our hope that this approach will be useful in providing a new understanding of treating a variety of fundamental problem in quantum computation analytically. \newline

The quantization approach adopted is Isham’s canonical group quantization (CGQ) \cite{Isham1}, a variant of geometric quantization \cite{Woodhouse} that place emphasis on the symmetries of the phase space, assuming a natural polarization. The CGQ is a global verison of canonical quantization that keeping close to the traditionl roots of quantum theory \cite{Ali}. Essentially, this program has been applied to the problem in quantum gravity \cite{Isham2,Isham3}, and particles and strings on tori \cite{Isham4}. Then it was applied to a quantum-mechanical system of particle on a torus in a constant magnetic field \cite{Zainuddin}, particle and Quantum Hall effect \cite{Bouketir}, particle of non-commutative configuration space \cite{Umar}, particle on the sphere in the presence of a magnetic monopole \cite{Reyes,Silva}, and nonperturbative quantum gravity \cite{Silva2}.\newline

The main idea of the CGQ is to treat classical phase spaces that are symplectic manifolds equipped with symplectic form $\omega$, and then identify a Lie group that respects the phase-space global kinematical symmetries of $\omega$ to be an ``appropriate'' canonical group $\mathcal{G}$. There are three type of Lie algebras available on the phase space (symplectic manifold)  that must be corresponded to each other prior to establish the quantization process. They are i) abstract algebra of the canonical group ii) commutator algebra of (global) Hamiltonian vector fields, and iii) Poisson bracket of classical observables. An irreducible unitary representation of $\mathcal{G}$ then gives a possible quantization of the system where inequivalent representations are considered to be the different physical realizations of quantum systems.\newline

The paper is organized as follows. In Section 2, we introduce the mathematical formulation of qubit geometry by identifying the sphere $S^2$ as a non-cotangent bundle phase space and describing its associated symplectic structure. Section 3 develops the holomorphic quantization procedure through the CGQ, emphasizing the role of holomorphic polarization in constructing wavefunctions via sections of a complex line bundle over $\mathbb{C}\text{P}^{1}$. In Section 4, we represent single-qubit quantum gates as M\"obius transformations, examining their geometric action on holomorphic wavefunctions and demonstrating their consistency with standard quantum gate operations. Finally, Section 5 summarizes our findings and outlines future directions, including generalizations to higher-dimensional systems and potential applications in quantum information theory.

\section{Mathematical formulation and non-cotangent bundle phase space $S^2$}
Before discussing holomorphic quantization, we provide a brief overview of the CGQ employed in our approach. The procedure begins by identifying a globally well-defined minimal set of preferred classical observables\footnote{It is analogous to the position, $q$ and momentum, $p$, observables in standard quantum mechanics on the configuration space $Q=\mathbb{R}^n$.} for a phase space $\mathcal{S}$ that is not necessarily a cotangent bundle. These observables generate all other classical observables and define a set of Hamiltonian vector fields on the phase space. Given an observable $f\in C^{\infty}(\mathcal{S},\mathbb{R})$, the corresponding Hamiltonian vector field $\xi_f$ is determined by
\begin{equation}\label{e:hamvec}
\xi_{f}\lrcorner\omega=-df.
\end{equation}
We assume that the commutator algebra of these vector fields is (anti)-homomorphic to the Poisson-brackets algebra of the observables. One then exponentiates these vector fields to generate the required canonical group $\mathcal{G}$. The final step is to find all inequivalent irreducible unitary representations of $\mathcal{G}$, producing inequivalent quantizations of the system. The self-adjoint generators in a unitary representation of $\mathcal{G}$ then produce the holomorphic part of the constructed vector fields.\newline

Our new approach differs from the original idea of the CGQ scheme (for a summary, see \cite{Isham1}). Instead of starting with a cotangent bundle phase space endowed with a configuration space $Q$, we work with a non-cotangent bundle (compact) phase space i.e. $\mathbb{C}\text{P}^{1}$ (that is constructed from $S^2$), which naturally leads to discrete spectra for quantum spin systems. Furthermore, the global kinematical symmetry is analyzed by constructing the canonical group $\mathcal{G}$ in a manner that does not take the semidirect product form. In particular, the Poisson bracket algebra of preferred classical observables and commutator algebra of Hamiltonian vector fields can be constructed to be (anti-)homomorphic to the algebra of the canonical group describing kinematic symmetries of $\mathbb{C}\text{P}^{1}$, just like the case of $S^2$. As a result, Mackey’s induced representation techniques no longer apply directly. Instead, the induced representation is obtained from the action of $\mathcal{G}$ on the space of holomorphic wavefunctions (section of line bundle), leading to possible quantizations.\newline

In standard quantum mechanics (QM), spin is interpreted as intrinsic angular momentum rather than arising from rotational or ``spinning'' motion. Meanwhile, spin-$\frac{1}{2}$ (or two-level quantum systems) carry the smallest unit of QIP. However, the classical phase space for a spin-$\frac{1}{2}$(or qubit) is not well-established. The minimal viable choice is the sphere $S^2$, since the simplest compact space $S^1$ is one-dimensional and cannot independently define a phase space.
We therefore take $S^2$ as a compact phase space $\mathcal{S}$ that is simply connected \cite{Andrews}, given by
\begin{equation}
	S^2=\{\vec{x}:\sum_{j=1}^{3}x^jx^j=1\}.
\end{equation}
This phase space is equipped with the natural symplectic form
\begin{equation}\label{Symplecticform}
	\omega=\sin\theta d\theta\wedge d\phi,
\end{equation}
where $0\leq \theta \leq \pi;0\leq \phi \leq  2\pi$. It is closed ($d\omega=0$) and is non-degenerate. The preferred set of observables is given by
\begin{equation}\label{ClassicalObservables}
		x_{1} = \sin\theta\cos\phi;\quad
		x_{2} = \sin\theta\sin\phi;\quad
		x_{3} = \cos\theta, 
	\end{equation}
where $0\leq\theta\leq\pi$; $0\leq\phi\leq 2\pi$.
Since $\theta$ and $\phi$, are periodic, they are not globally well-defined continuous functions on $S^2$. Instead, we use the observables in (\ref{ClassicalObservables}) as a minimal globally well-defined set to ensure closure of the Poisson bracket algebra. \newline

Geometrically, $S^2$ can be viewed as the homogeneous space $SO(3)/SO(2)$, suggesting $SO(3)$ as the natural choice for the canonical group $\mathcal{G}$. From the fundamental equation (\ref{e:hamvec}) the corresponding Hamiltonian vector fields (HamVF($\mathcal{S}$)) associated with the observables (\ref{ClassicalObservables}) are
\begin{align}\label{SetofHamVF}
	\xi_{1}&= \sin\phi\frac{\partial}{\partial\theta}+\cot\theta\cos\phi\frac{\partial}{\partial\phi},\quad 
	\xi_{2}&= -\cos\phi\frac{\partial}{\partial\theta}+\cot\theta\sin\phi\frac{\partial}{\partial\phi},\quad
	\xi_{3}&= -\frac{\partial}{\partial\phi},
\end{align}
These are precisely the standard angular momentum operators. Their commutator algebra satisfies
 \begin{equation}\label{COMM}
	[\xi_{j},\xi_{k}] = \varepsilon_{jkl}\xi_{l},
\end{equation}
where $\varepsilon_{jkl}$ is the totally antisymmetric cyclic permutation, that corresponds to the Lie algebra $\mathfrak{so(3)}$. The Poisson bracket algebra of the global position observables (\ref{ClassicalObservables}) is given by
\begin{equation}\label{PB1}
\{x_{j},x_{k}\}=\omega (\xi_{j},\xi_{k})=-\varepsilon_{jkl}x_{l}.
\end{equation} 
These structures reveal an (anti-)homomorphism to the abstract algebra of $SO(3)$, further justifying the choice of $SO(3)$ as the canonical group.\newline

An alternate route is to obtain the transitive action of the group $\mathcal{G}$ on the phase space $(\mathcal{S},\omega)$  given by
\begin{equation}\label{OPSSO2}
	(g,\vec{x}) \longmapsto \ell_{g}(\vec{x})= \vec{x}',
\end{equation}
where $g:=\exp(-iJ)\in SO(3)$ is constructed from the exponential map of Lie algebra  $J\in \mathfrak{so(3)})$ as one-parameter subgroups (OPS) elements, and  $\vec{x}\in S^{2}$. These transformations generate the same HamVF$(\mathcal{S})$ ( \ref{SetofHamVF}) through their integral curves on phase space. \newline

If the quantum wavefunction were not required to depend on only half of the phase space coordinates, then $SO(3)$ might serve as a suitable canonical group for quantizing $S^2$. However, a fundamental requirement is the existence of a natural polarization of the phase space \cite{Woodhouse}. The sphere $S^2$ lacks such a polarization due to its nonvanishing Euler class \cite{Vaisman}. Thus, the need for a compact phase space $\mathbb{C}\text{P}^1$ to impose the holomorphic polarization in CGQ.


\section{Holomorphic quantization} 
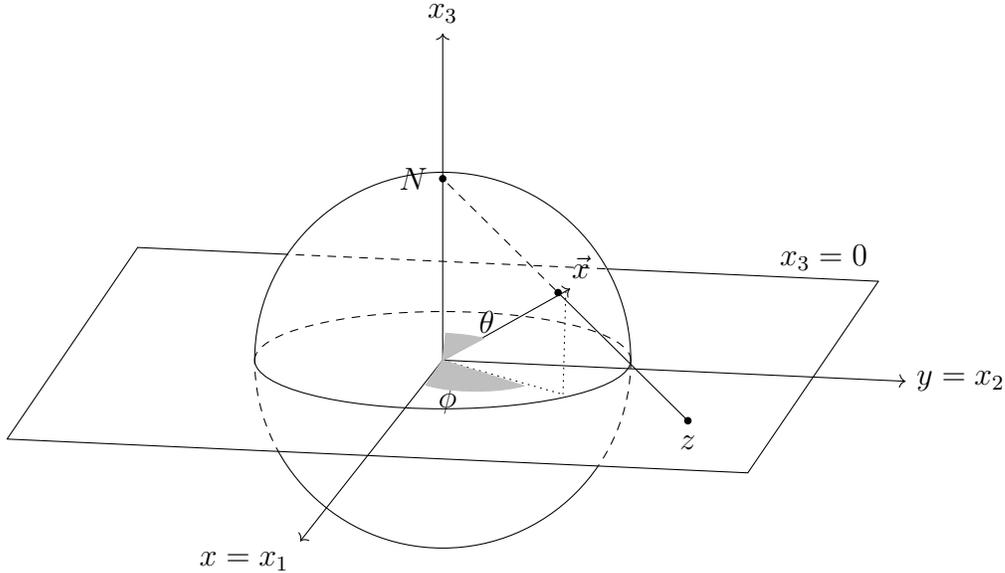
\begin{figure}
	\centering
	\begin{tikzpicture}[declare function={%
			stereox(\x,\y)=2*\x/(1+\x*\x+\y*\y);%
			stereoy(\x,\y)=2*\y/(1+\x*\x+\y*\y);%
			stereoz(\x,\y)=(-1+\x*\x+\y*\y)/(1+\x*\x+\y*\y);},scale=2.5,
		line join=round,line cap=round,
		dot/.style={circle,fill,inner sep=1pt}]
		\path[save path=\pathSphere] (0,0) circle[radius=1];
		\begin{scope}[3d view={\myaz}{15}]
			\draw (-2,2) -- (-2,-2) coordinate (bl) -- (2,-2) coordinate (br)-- (2,2)
			node[above left]{$x_{3}=0$};
			\begin{scope}
				\tikzset{protect=\pathSphere}
				\draw (-2,2) -- (2,2);
			\end{scope}
			\begin{scope}
				\draw[->] (0,0)--(2.5,0) node[right] {$y=x_{2}$};
				\draw[->] (0,0)--(0,0,1.8) node[above] {$x_{3}$};
				\draw[->] (0,0)--(-.100,-3.8) node[below left] {$x=x_{1}$};
			\end{scope}
			\begin{scope}
				\clip[reuse path=\pathSphere];
				\draw[dashed] (-2,2) -- (2,2);
			\end{scope}
			\begin{scope}[canvas is xy plane at z=0]
				\draw[dashed] (\myaz:1) arc[start angle=\myaz,end angle=\myaz+180,radius=1];
				\draw (\myaz:1) arc[start angle=\myaz,end angle=\myaz-180,radius=1];
				\path[save path=\pathPlane] (\myaz:2) -- (\myaz+180:2) --(bl) -- (br) -- cycle;
				\begin{scope}
					\clip[use path=\pathPlane];
					\draw[dashed,use path=\pathSphere];
				\end{scope}
				
				\begin{scope} 
					\tikzset{protect=\pathPlane}
					\draw[use path=\pathSphere];
				\end{scope}0.5
			\end{scope}
			\draw[->] (0,0)--(.6,0.5,0.29);
			\draw [dotted] (0.40,1.5)-- (0.75,-0.57);
			\draw [dotted] (0,0)-- (0.75,-0.58);
			\draw (-0.03,-0.3) node[anchor=north west] {$\phi$};
			\draw (0.05,0.5,0.2) node[anchor=north west] {$\theta$};
			\draw [shift={(0,0)}, lightgray, fill, fill opacity=0.1] (0,0) -- (73.9:0.50) arc (56.7:90.:0.4) -- cycle;
			\draw [shift={(0,0)}, lightgray, fill, fill opacity=0.1] (0,0) -- (-90:0.52) arc (-130:-33.2:0.35) -- cycle;
			\draw (1.5,-1,0) node[dot,label=below:{$z$}]{}
			-- ({stereox(1.5,-1)},{stereoy(1.5,-1)},{stereoz(1.5,-1)})
			node[dot,label=above right:{$\vec{x}$}](I){};
			\draw[dashed] (I) -- (0,0,1) node[dot,label=left:{$N$}]{}; 
		\end{scope}
	\end{tikzpicture}
	\captionof{figure}{Stereographic projection from the north pole}\label{stereoRiemann}
\end{figure}
It is well-known that $S^2$ is homeomorphic to $\mathbb{C}\text{P}^1$ via stereographic projection \cite{Jones} as illustrated in Figure \ref{stereoRiemann}. This projection maps a point $\vec{x} = (x^1, x^2, x^3)$ on $S^2$ to a point $z=x+iy$ on the extended complex plane $\hat{C}:=\mathbb{C}\cup \{\infty\}$. Specifically, the line passing through the north pole $N = (0, 0, 1)$ and the point $\vec{x}$ on $S^2$ intersects $\hat{C}$ at $z=\cot(\frac{\theta}{2})e^{i\phi}=\infty$. This construction explicitly identifies $S^2$ with the Riemann sphere $\mathbb{C}\text{P}^1$, a natural setting for holomorphic quantization \cite{Woodhouse}. The symplectic form on the resultant phase space
$\mathcal{\tilde{S}}=\mathbb{C}\text{P}^{1}$ is
\begin{equation}\label{SympFormCP1}
	\Omega= \frac{2idz\wedge d\bar{z}}{(1+z\bar{z})^{2}},
\end{equation}
and the complex canonical observables are given by
\begin{equation}\label{CCU}
	\begin{split}
			u_{1} = \frac{z+\bar{z}}{(z\bar{z}+1)};\quad
			u_{2} =-i\frac{z-\bar{z}}{(z\bar{z}+1)};\quad
			u_{3} = \frac{z\bar{z}-1}{(z\bar{z}+1)}.
		\end{split}
\end{equation}
Unlike $S^2$, the Riemann sphere admits a natural polarization between $z$ and $\overline{z}$, facilitating holomorphic quantization. The symplectic form $\Omega$ encodes an inherent extra hidden discrete symmetry under complex conjugation, $z\longmapsto \overline{z}$. Incorporating this symmetry leads to the natural choice of the canonical group as the double cover $SU(2)$ of the $SO(3)$ symmetry of $S^2$.\newline

Furthermore, $\mathbb{C}\text{P}^1$ is linked to $\mathbb{C}$ through homogeneous coordinates $(z_1,z_2)$, where the inhomogeneous coordinate is given by $z=\frac{z_1}{z_2}$. This identification aligns with the standard representation of qubits in $\mathbb{C}^2$. The stereographic projection from the north and south poles naturally defines two coordinate charts on $SU(2)$, corresponding to the inhomogeneous coordinates $z$ and $\frac{1}{z}\equiv \overline{z}$ , satisfying $\overline{z}z=1$. This two-chart structure underlies the transition functions essential for a consistent geometric quantization framework.\newline

By treating the inhomogeneous coordinates as equivalence classes i.e. $(z_1,z_2)^{T}\equiv (z,1)^{T}$, one observes that the $SU(2)$-action $\ell_{\tilde{g}}$ on $\mathbb{C}\text{P}^{1}$ is of M\"obius transformations 
\begin{equation}\label{FLTAction}
		 \begin{pmatrix} \alpha && \beta \\ 
			-\bar{\beta}     &&\bar{\alpha}\end{pmatrix} \begin{pmatrix} z \\ 
			1 \end{pmatrix} \longmapsto  \begin{pmatrix} \alpha z+ \beta\\
			-\overline{\beta} z+ \overline{\alpha} \end{pmatrix} \equiv \begin{pmatrix} \frac{\alpha z+ \beta}{-\overline{\beta} z+ \overline{\alpha}}\\ 
			1 \end{pmatrix};~\alpha, \beta \in \mathbb{ C},
\end{equation}
where $\alpha\overline{\alpha}+\beta\overline{\beta}=1$ and $-\overline{\beta} z+ \overline{\alpha}\neq0$, integrating a nontrivial phase factors. The Hamiltonian vector fields of $\tilde{S}$ (HamVF($\tilde{S}$)) can then be constructed through the similar method as $S^2$ to obtain 
\begin{equation}\label{HamiltonianVFldCP1}
	\tilde{\xi}_{1}=\frac{i}{2}(1-z^{2})\frac{\partial}{\partial z}-\frac{i}{2}(1-\bar{z}^{2})\frac{\partial}{\partial\bar{z}},\quad
	\tilde{\xi}_{2}= \frac{1}{2}(1+z^{2})\frac{\partial}{\partial z}+\frac{1}{2}(1+\bar{z}^{2})\frac{\partial}{\partial \bar{z}},\quad
	\tilde{\xi}_{3}= iz\frac{\partial}{\partial z}-i\bar{z}\frac{\partial}{\partial\bar{z}}.
\end{equation}
Alternatively, (\ref{HamiltonianVFldCP1}) can also be constructed by substituting the OPS elements $\tilde{g}:=\exp(-\frac{i\theta\sigma}{2})$ where $\sigma=\{\sigma_j\}_{j=1}^{3}\in \mathfrak{su(2)}$ are Pauli matrices, into (\ref{FLTAction}). The commutator algebra of (\ref{HamiltonianVFldCP1}) is equivalent to (\ref{COMM}) and corresponds to the Lie algebra $\mathfrak{su(2)}$ meanwhile the Poisson bracket algebra of observables (\ref{CCU}) is equivalent to (\ref{PB1}). In fact, at the face value, locally their algebraic structures are isomorphic.\newline

Choosing the natural holomorphic polarization, one can construct inequivalent irreducible unitary representations of the canonical group $SU(2)$ through the action on the representation space for $\mathcal{\tilde{G}}$ which its elements are holomorphic wavefunctions (sections of the line bundle). We consider first the nontrivial Hopf bundle,
\begin{equation}\label{HOPF}
	U(1)\simeq S^1\longrightarrow S^{3}\longrightarrow S^{2}\simeq \mathbb{ C}\text{P}^1,
\end{equation} 
as our principal $U(1)$-bundle (with the first Chern number $\frac{i}{2\pi}\int_{\mathbb{C}\text{P}^1}\Omega=2\pi$) then we define an associated vector bundle\footnote{Here, the UIR of $U(1)$ is indexed by its character \textit{i.e.} $\chi(\lambda)= \lambda^{n}; n\in\mathbb{Z}$ and such an integer $n$ topologically is related to the first Chern number.} over $ \mathbb{ C}\text{P}^1$,
\begin{equation}
	\mathbb{ C}\longrightarrow \mathcal{L}:=S^3\times_{U(1)}\mathbb{C}\longrightarrow \mathbb{ C}\text{P}^1,
\end{equation}
where for all $\lambda\in U(1)$ there is an equivalence relation $(w\lambda, \mathcal{U}(\lambda^{-1})v) \sim (w=(w_1,w_2),v)$ that gives an equivalence class $[w,v]$ to be the element of $\mathcal{L}$ (for a more details idea of this bundle one can refer \cite{Isham1}). The space of local sections of $\mathcal{L}$,
\begin{equation}\label{SpaceofSect2}
	\Gamma_{\text{hol}}(U\times \mathbb{C})\simeq \{\psi:U\subset\mathbb{C}\text{P}^1\longrightarrow \mathbb{C}\big| \psi(z)= \mathcal{U}(\lambda^{-1})\psi(z)\}, 
    \end{equation}
will be the representation space for $\mathcal{\tilde{G}}$. Since we have the M\"obius action in (\ref{FLTAction}), this action has a nontrivial lift $\ell_{\tilde{g}}^{\uparrow}$ to the $\mathcal{L}$, as shown in the following commutative diagram
\begin{align*}
	\xymatrixrowsep{0.4in}
	\xymatrixcolsep{0.4in}
	\xymatrix{	
\mathbb{ C}\ar[r]			&\mathcal{L}\ar_{\pi_{\mathbb{ C}}}[d]\ar[r]^{\ell_{\tilde{g}}^{\uparrow}}& \mathcal{L} \ar^{\pi_{\mathbb{ C}}}[d]\\
		&\mathbb{ C}\text{P}^{1}\ar[r]^{\ell_{\tilde{g}}}&\mathbb{ C}\text{P}^{1}} 
\end{align*}
such that $\ell_{\tilde{g}}\cdot\pi_{\mathbb{ C}}= \pi_{\mathbb{ C}}\cdot \ell_{\tilde{g}}^{\uparrow}$. Here, the $\ell_{\tilde{g}}^{\uparrow}$-action give rise to an action of the $U(1)$ structure group of $\mathcal{L}$ on the fibers $v$ in terms of complex variables, 
\begin{equation}\label{EQ447}
			\ell^{\uparrow}_{\tilde{g}} \Big([w,v]\Big) \longmapsto \Big[(w'_{1},w'_{2}),\Big(\frac{\beta w+\overline{\alpha}}{|\beta w+\overline{\alpha}|}\Big)\cdot v\Big)\Big],
	\end{equation}
    	where $w'_{1}:= \alpha w_{1} + \beta w_{2} , w'_{2}:= -\overline{\beta} w_{1}+ \overline{\alpha}w_{2} \in S^{3};\tilde{g}\in SU(2)$.
In the following argument, the structure group of the fibers $v$ will appear as the phase factor of the wavefunctions in the representation operators.\newline

Thus, through the $\ell^{\uparrow}_{\tilde{g}}$-action on the space of local sections (\ref{SpaceofSect2}) one has the following representation of $\tilde{\mathcal{G}}$ 
\begin{equation}\label{GenF}
	(\mathcal{U}_{\tilde{g}}\psi)(z) = ( \beta z+\overline{\alpha} )\psi\Big(\frac{\alpha z- \overline{\beta}}{\beta z+\overline{\alpha}} \Big),
\end{equation}
where $\psi(z)=\sum_{j=-l}^{l}c_{j}z^{l+j} (c_j$ are complex coefficients) is the holomorphic wavefunction\footnote{It is constructed from the local trivialization $\pi^{-1}(U_1)\simeq U_1\times \mathbb{C}$ of $\mathcal{L}$.} and (\ref{GenF}) is the simplest form of homogenized polynomial. By substituting $\tilde{g}:=\exp(-\frac{i\theta\sigma}{2})$ into (\ref{GenF}) one gets the following representation of $\mathcal{\tilde{G}}$,
	\begin{align}\label{opsa}
		(\mathcal{U}_{\tilde{g}_{1}}\psi)(z)&=(i\sin\frac{\theta_{1}}{2} z+ \cos \frac{\theta_{1}}{2})\psi\Big(\frac{ \cos \frac{\theta_{1}}{2}z +i\sin\frac{\theta_{1}}{2}}{i\sin\frac{\theta_{1}}{2} z+ \cos \frac{\theta_{1}}{2}}\Big),\\\label{OPSb}
		(\mathcal{U}_{\tilde{g}_{2}}\psi)(z)&=(\sin\frac{\theta_{2}}{2} z+ \cos \frac{\theta_{2}}{2})\psi\Big(\frac{ \cos \frac{\theta_{2}}{2}z -\sin\frac{\theta_{2}}{2}}{\sin\frac{\theta_{2}}{2} z+ \cos \frac{\theta_{2}}{2}}\Big),\\\label{OPS3c}
	(\mathcal{U}_{\tilde{g}_{3}}\psi)(z)&=e^{-\frac{i\theta_{3}}{2}}\psi(e^{i\theta_{3}}z),
	\end{align}
and a set of ``spin angular momentum'' operators is constructed, that is the holomorphic part of the constructed vector fields (\ref{HamiltonianVFldCP1}) (after polarization) for the system appended with a connection-type term, as follows,
	\begin{align}\label{AngMomOperator}
	 \hat{S}_{1} = \frac{z}{2}+\frac{1}{2}(1-z^{2})\frac{\partial}{\partial z}, 
		\hat{S}_{2} = -\frac{iz}{2}+\frac{i}{2}(1+z^{2})\frac{\partial}{\partial z}, 
		\hat{S}_{3} = -\frac{1}{2}+z\frac{\partial}{\partial z}.
	\end{align}
Its Casimir operator is  $\hat{S}^{2}=\hat{S}^{2}_{1}+\hat{S}^{2}_{2}+\hat{S}^{2}_{3}=\frac{3}{4}\hat{\mathbb{I}}$, and the eigenfunctions of $\hat{S}_{3}$ are $c_{-\frac{1}{2}}$ and $c_{\frac{1}{2}}z$. The eq. (\ref{AngMomOperator}) obey the commutation relation 
\begin{equation}\label{COMM2}
    [\hat{S}_j,\hat{S}_k]=i\varepsilon_{jkl}\hat{S}_{l}.
\end{equation}
Furthermore, one can generalize the representation by noting that the Hopf bundle (\ref{HOPF}) being nontrivial (first Chern number $2\pi n;n\in \mathbb{Z}$). Thus the representation (\ref{GenF}) becomes
\begin{equation}\label{GenF2}
	(\mathcal{U}^l_{\tilde{g}}\psi)(z) = ( \beta z+\overline{\alpha} )^{2l}\psi\Big(\frac{\alpha z- \overline{\beta}}{\beta z+\overline{\alpha}} \Big),
\end{equation}
and the general holomorphic part of the constructed vector fields after polarization also serves in part as general spin angular momentum operators for the system associated with a general connection-type term $l$,
\begin{align}\label{AngMomop1}
	 \hat{S}_{1} = lz+\frac{1}{2}(1-z^{2})\frac{\partial}{\partial z},\quad 
	\hat{S}_{2} = -ilz+\frac{i}{2}(1+z^{2})\frac{\partial}{\partial z},\quad 
	\hat{S}_{3} = -l+z\frac{\partial}{\partial z}, 
\end{align}
obeying a similar commutation relation as (\ref{COMM2}). Thereafter, one could classify the wavefunctions as in standard $SU(2)$ eigenfunctions with the help of a Casimir operator $l(l+1)$. Throughout (\ref{AngMomop1}), the ladder operators are obtained 
\begin{equation}\label{LAD11}
	\hat{S}_{+} = 2lz-z^{2}\frac{\partial}{\partial z},\quad
	\hat{S}_{-} = \frac{\partial}{\partial z};\quad 
\end{equation}
satisfying 
\begin{equation}\label{LAD}
		[\hat{S}_{+}, \hat{S}_{-}] =  2\hat{S}_{3},\quad
		[\hat{S}_{3}, \hat{S}_{\pm}] =\pm\hat{S}_{\pm}. 
\end{equation} 
In each case, the eigenfunction of $\hat{S}_3$ is a monomial $z^{l+j}$ of $\psi(z)=\sum_{j=-l}^{l}c_{j}z^{l+j}$, and can use standard methods \cite{PhDAHAS} to generate representations of $SU(2)$ that will be discussed shortly after this. Let us write the monomial $ z^{l+j}\equiv \psi^{l}_{j}$ (that will be used interchangeably after this), thus the action of the $\hat{S}$'s on the monomial $\psi^{l}_{j}$ produces $(l\mp j)\psi^{l}_{j\pm 1}$, therefore  
\begin{align}
	\hat{S}_{\pm}\psi^{l}_{j}=\sqrt{(l\pm j+1)(l\mp j)}\psi^{l}_{j\pm 1},\quad \hat{S}_{3}\psi^{l}_{j}=j\psi^{l}_{j},
\end{align}
where $\psi^{l}_{j}$ is the canonical basis, which treats the raising and lowering operators in a manifestly similar way to the standard angular operators in QM. Therefore, from all the above constructions it is straightforward to deduce that the eigenvalue $j$ is generally, $$j\in \{-l,-l+1,...,l-1,l\};~l:=\frac{n}{2};n\in \mathbb{Z}.$$ To complete the quantization, we wish to introduce a Hermitian vector bundle (which is an inner product as in standard QM). Choose
\begin{equation}\label{ONbasis}
		\psi^{l}_{j} =\frac{z^{l+j}}{\sqrt{(l+j)!(l-j)!}};\quad -l\leq j\leq l,
\end{equation}
as an orthonormal basis, thus for any holomorphic wavefunctions in the Hilbert space, the inner product can be written as the holomorphic integral 
 \begin{equation}\label{INNERPRODUCT}
	\frac{i} {2\pi}\sum_{j=-l}^{l}\int_{\mathbb{C}\text{P}^{1}}\bar{c_j}c_j\frac{(l+j)!(l-j)!}{2l!} \Omega_{(n)}, 
\end{equation}
where $\Omega_{(n)}=\frac{2in\,dz\wedge d\bar{z}}{(1+z\bar{z})^{2}}$ is the integration over $\mathbb{C}\text{P}^{1}$ as the Hilbert space measure.\newline

From (\ref{ONbasis}), we shall find matrix entries
\begin{equation}\label{REL}
	\mathcal{U}^{l}_{kj}(\tilde{g})=\braket{e_{k},\mathcal{U}^{l}_{\tilde{g}}e_j},
\end{equation}
where $	e_j$ is a basis that is equivalent to $\psi^{l}_{j}$. Notice that the inner product (\ref{INNERPRODUCT}) can be expressed by means of the differential operators \begin{equation}\label{DIF}
	D^{l}_{k}=\sum^{l}_{k=-l} (l-k)!c_k \Big(\frac{d}{d z}\Big)^{l+k}\Big|_{z=0};\quad -l\leq k \leq l
\end{equation}
associated with the monomial $z^{l+j}$. The inner product is given by
\begin{equation*}\label{NormIn}
	\Braket{D^{l}_k,\psi^{l}_{j}}\Big|_{z=0},
\end{equation*}
and by using this new form of inner product, the $kj$-th matrix element (the explicit derivation can be referred to \cite{PhDAHAS}) is given in terms of Euler angles as a product of an exponential function, trigonometric terms and Jacobi polynomial,
\begin{equation}\label{MAT22}
	\begin{split}
		\mathcal{U}^{l}_{kj}(\theta_3,\theta_2,\theta'_3)= \frac{(-1)^{l+k}}{2^{l}}&\sqrt{\frac{(l+k)!}{(l-k)!(l-j)!(l+j)!}}e^{i(j\theta_3 + k\theta'_3)}\\&\times(\sin\frac{\theta_2}{2})^{j-k}(\cos\frac{\theta_2}{2})^{-(k+j)}P^{(j-k;-(k+j))}_{l+k}(\cos\theta_2).
	\end{split}
\end{equation}
where $P^{(j-k;-(k+j))}_{l+k}(\cos\theta_2)$ is equivalent to the Rodrigues' formula \cite{Gurarie}. Therefore, equation (\ref{MAT22}) represents the unitary operator in terms of numerical functions compare to (\ref{GenF2}). However, it is interesting to discuss the wavefunctions of qubits operations through the latter.
\section{Single-Qubit Gates in the Holomorphic Formalism}

Next, we determine the representation corresponding to the single-qubit unitary operation
using (\ref{opsa}) - (\ref{OPS3c}). As an example the construction of Hadamard gate $H$, from OPS elements $\{\tilde{g}_1,\tilde{g}_2,\tilde{g}_3\}$ of $\tilde{g}:=\exp(-i\frac{\theta\sigma}{2})\in SU(2)$ it produces
\begin{equation}\label{Hadamard}
\tilde{g}_{12}:=\tilde{g}_{H}=\begin{pmatrix}
C_2C_1+iS_2S_1   && iC_2S_1+S_2C_1\\
\\
-S_2C_1 + iC_2S_1  &&  -iS_2S_1+C_2C_1  
\end{pmatrix},
\end{equation} 
with $C_j\equiv \cos(\frac{\theta_j}{2})$ and $S_j\equiv \sin(\frac{\theta_j}{2});j=1,2$. By substituting (\ref{Hadamard}) into (\ref{GenF}) and taking $\theta_2=\frac{\pi}{2}$ and $\theta_1=\pi$, we obtain the representation 
\begin{equation}\label{UIRHad}
	(\mathcal{U}_{\tilde{g}_{H}}\psi)(z)=(z-1)\psi\Big(\frac{z+ 1}{z-1}\Big)\simeq \psi\Big(\frac{z+ 1}{z-1}\Big),
\end{equation}
of $H$ on $\mathbb{ C}\text{P}^1$ where for $z\longrightarrow \frac{z+ 1}{z-1}$ taking $\theta=\frac{\pi}{4},\phi=\{0,\pi\}$ gives fixed points $\{1\pm \sqrt{2}\}$. These transformations act on the basis states $\ket{0}$ and $\ket{1}$ is represented by the stereographic coordinates $z$: $\ket{0}$ corresponds to $z=\infty$ and $\ket{1}$ corresponds to $z=0$. Explicitly, for (\ref{UIRHad}) if $z=\infty$, we obtain 
$$\quad \psi\Big(\frac{z+ 1}{z-1}\Big)= \psi\Big(\frac{\infty+ 1}{\infty-1}\Big)=1,$$ 
which corresponding to $\ket{0}\longmapsto \frac{\ket{0}+ \ket{1}}{\sqrt{2}}$, and if $z= 0$, we obtain 
$$\quad \psi\Big(\frac{z+ 1}{z-1}\Big)= \psi\Big(\frac{0+ 1}{0-1}\Big)=-1,$$ 
which corresponding to $\ket{1}\longmapsto \frac{\ket{0}- \ket{1}}{\sqrt{2}}$. Moreover, the fixed points align with the eigenstates of the corresponding standard Hadamard logic gate i.e recall
\[
H = \frac{1}{\sqrt{2}} \begin{bmatrix}
1 & 1 \\
1 & -1
\end{bmatrix}.
\]
To find its eigenstates, one solves the eigenvalue equation
 $H\ket{\psi}=\lambda \ket{\psi} $
and yields two normalized eigenstates corresponding to eigenvalues $\lambda = \pm 1$:
\[
|\psi_{+}\rangle = \frac{1}{\sqrt{4 + 2\sqrt{2}}} \begin{bmatrix} 1 + \sqrt{2} \\ 1 \end{bmatrix}, \quad
|\psi_{-}\rangle = \frac{1}{\sqrt{4 - 2\sqrt{2}}} \begin{bmatrix} 1 - \sqrt{2} \\ 1 \end{bmatrix}.
\]
These eigenstates lie on irrational points of the Riemann sphere and are fixed under M\"obius transformations associated with $H$.\newline 

Furthermore, the wavefunction of the phase shift gate that shifts the phase of the $\ket{0}$ state relative to the $\ket{1}$ state, as in a common single-qubit formalism that modifies the phase of the quantum state, can be constructed as follows. Recall (\ref{OPS3c}) and set $\theta_3=\pi$ thus it yields the $S$-gate wavefunction
\begin{equation}
    (\mathcal{U}_{\tilde{g}_{3}}\psi)(z)=\psi(iz),
\end{equation}
where the $z\mapsto \psi(iz)$ fixing $z=0$ that is a $\pi$ radian rotation around $z$-axis on $\mathbb{C}\text{P}^1.$ This is corresponding to the common $S$-gate on the Bloch sphere. And if we set $\theta_3=\frac{\pi}{2}$ thus it yields the wavefunction of $T$-gate as follows
\begin{equation}
    (\mathcal{U}_{\tilde{g}_{3}}\psi)(z)=\frac{1}{\sqrt{2}}\psi(z+iz).
\end{equation}
that is $\frac{\pi}{4}$ radian rotation on $z$-axis fixing $z=0.$ Geometrically, this is equivalent to tracing a horizontal fibre (or circle) of the total space over $\mathbb{C}\text{P}^1$ along the $z$-axis by $\theta_3$. \newline

In conclusion, the $S$- and $T$-gates only fix $0$, meaning they act as pure (fibre) phase shifts. The Hadamard gate is the most nontrivial, fixing two intermediate points, and the Pauli gates are simple reflections, with clear fixed points at geometrically well-defined locations. All of these results confirm that M\"obius transformations on $\mathbb{C}\text{P}^1$ correctly encode single-qubit gates, and the representation remains valid within the holomorphic wavefunction framework constructed via sections of the line bundle over $\mathbb{C}\text{P}^1$. Table \ref{REPTAB} summarizes the correspondence between the common gates and our gates' holomorphic wavefunction, and Figure \ref{FIG1} shows the fixed points of quantum gate wavefunctions on $\mathbb{C}\text{P}^1$.\newline

These results provide a novel perspective on qubit geometry using the wave mechanics formalism derived from Isham’s canonical group quantization on a non-cotangent bundle phase space. The Möbius action on holomorphic wavefunctions naturally encodes single-qubit operations, yielding representations that align with standard quantum gates while offering a geometric interpretation of quantum transformations. This framework confirms the validity of standard quantum gates in a holomorphic setting and highlights the role of complex analytic structures in quantum computation.

\begin{center}
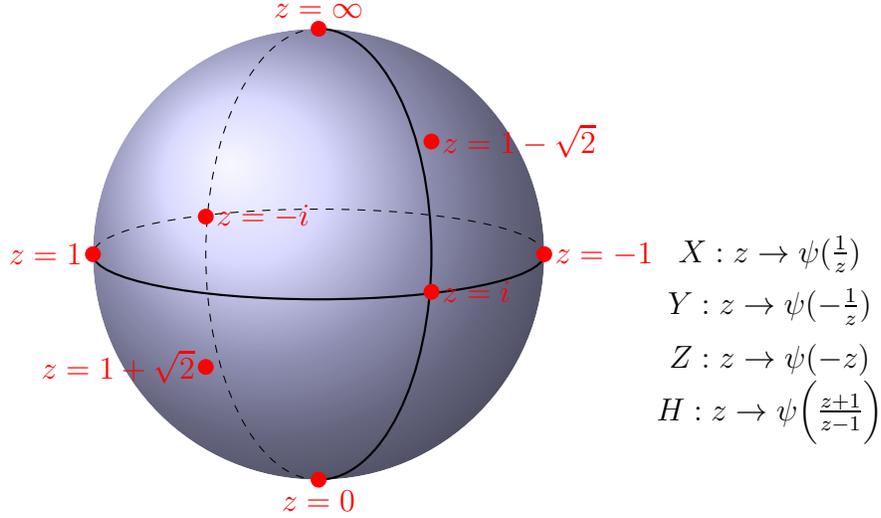
\begin{figure}
    \begin{tikzpicture}
        \shade[ball color=blue!20] (0,0) circle (3cm);
        
        \draw[thick] (-3,0) arc (180:360:3cm and 0.6cm);
        \draw[dashed] (3,0) arc (0:180:3cm and 0.6cm);
        
        \draw[thick] (0,-3) arc (-90:90:1.5cm and 3cm);
        \draw[dashed] (0,3) arc (90:270:1.5cm and 3cm);
        
        \fill[red] (0,3) circle (3pt) node[above]{$z=\infty$};
        \fill[red] (0,-3) circle (3pt) node[below]{$z=0$};
        \fill[red] (3,0) circle (3pt) node[right]{$z=-1$};
        \fill[red] (-3,0) circle (3pt) node[left]{$z=1$};
        \fill[red] (1.5,1.5) circle (3pt) node[right]{$z=1-\sqrt{2}$};
        \fill[red] (-1.5,-1.5) circle (3pt) node[left]{$z=1+\sqrt{2}$};
        \fill[red] (-1.5,0.5) circle (3pt) node[right]{$z=-i$};
        \fill[red] (1.5,-0.5) circle (3pt) node[right]{$z=i$};
        
        \node at (6,0) { $X:z \to \psi(\frac{1}{z})$};
        \node at (6,-0.7) { $Y:z \to \psi(-\frac{1}{z})$};
        \node at (6,-1.4) { $Z: z \to \psi(-z)$};
        \node at (6,-2.1) { $H:z \to \psi\Big(\frac{z+1}{z-1}\Big)$};
    \end{tikzpicture}
    \caption{Fixed point of quantum gate wavefunction on $\mathbb{C}\text{P}^1$}\label{FIG1}
    \end{figure}
\end{center}
\begin{table}
\centering
\caption{Geometric interpretation of quantum gates via their holomorphic wavefunctions, including their action on the Riemann sphere, fixed points, and induced transformations.}
\label{REPTAB}
\begin{tabularx}{1\textwidth}{X  X X X X }
\toprule
\textbf{Common Gate} & \textbf{Symbol} & \textbf{Gates' Wavefunctions}&  \textbf{Fixed Points} & \textbf{Corresponding Eigenstates}\\
\midrule
$\begin{bmatrix} 1 & 0 \\ 0 & 1 \end{bmatrix}$ & \begin{quantikz} &\gate{I}& \end{quantikz} & $\psi(z)$ & All points on $\mathbb{C}\text{P}^1$& Every point on $\mathbb{C}\text{P}^1$ is an eigenstate of the $I$-gate \\
$\begin{bmatrix} 0 & 1 \\ 1 & 0 \end{bmatrix}$ & \begin{quantikz} &\gate{X}& \end{quantikz} & $\psi(\frac{1}{z})$ & $\{\pm 1\}$ &$\Big\{\begin{pmatrix}
		  1 \\ 
				1 
		\end{pmatrix}, \begin{pmatrix}
		  1 \\ 
				-1 
		\end{pmatrix}\Big\}$\\
$\begin{bmatrix}
		   0 &&-i\\
           i && 0
		\end{bmatrix}$ & \begin{quantikz} &\gate{Y}& \end{quantikz} & $\psi(-\frac{1}{z})$ & $\{\pm i\}$&$\Big\{\begin{pmatrix}
		  0 \\ 
				i 
		\end{pmatrix}, \begin{pmatrix}
		  -i \\ 
				0 
		\end{pmatrix}\Big\}$ \\
$\begin{bmatrix}
		   1  &&0\\
            0&&-1
		\end{bmatrix}$ & \begin{quantikz} &\gate{Z}& \end{quantikz} & $-\psi(z)$ & $\{\infty,0\}$&$\Big\{\begin{pmatrix}
		  1 \\ 
				0 
		\end{pmatrix}, \begin{pmatrix}
		  0 \\ 
				1 
		\end{pmatrix}\Big\}$ \\
$\frac{1}{\sqrt{2}}\begin{bmatrix}
		  1  &&1\\
           1 &&-1
		\end{bmatrix}$ & \begin{quantikz} &\gate{H}& \end{quantikz} & $\psi(\frac{z+1}{z-1})$ & $\{1\pm\sqrt{2}\}$&$\Big\{\begin{pmatrix}
		  1\pm \sqrt{2} \\ 
				1 
		\end{pmatrix}\Big\}$\\
$\begin{bmatrix}
		  1  &&0\\
           0 && i
		\end{bmatrix}$ &\begin{quantikz} &\gate{S}& \end{quantikz} & $\psi(iz)$ & $\{0\}$ & $\Big\{\begin{pmatrix}
		  1 \\ 
				0 
		\end{pmatrix}, \begin{pmatrix}
		  0 \\ 
				i 
		\end{pmatrix}\Big\}$ \\
$\begin{bmatrix}
		  1  &&0\\
           0 && e^{i\pi/4}
		\end{bmatrix}$ & \begin{quantikz} &\gate{T}& \end{quantikz} & $\frac{1}{\sqrt{2}}\psi(z+iz)$ & $\{0\}$ & $\Big\{\begin{pmatrix}
		  1 \\ 
				0 
		\end{pmatrix}, \begin{pmatrix}
		  0 \\ 
				e^{i\pi/4} 
		\end{pmatrix}\Big\}$\\
$\begin{bmatrix}
		  \cos\frac{\theta}{2}     & -i\sin\frac{\theta}{2}\\ 
				-i\sin\frac{\theta}{2}   & \cos\frac{\theta}{2}
		\end{bmatrix}$ & \begin{quantikz}& \gate{R_X(\theta)}& \end{quantikz} & $\psi\Big(\frac{ \cos \frac{\theta_1}{2}z +i\sin\frac{\theta_1}{2}}{i\sin\frac{\theta_1}{2} z+ \cos \frac{\theta_1}{2}}\Big)$ & $\{\pm 1\}$& Rotation about $X$-axis \& similar as $NOT(X)$-gate \\
$\begin{bmatrix}
		 \cos\frac{\theta}{2}    & -\sin\frac{\theta}{2}\\ 
				\sin\frac{\theta}{2}    & \cos\frac{\theta}{2} 
		\end{bmatrix}$ & \begin{quantikz} &\gate{R_Y(\theta)}& \end{quantikz} & $\psi\Big(\frac{ \cos \frac{\theta_2}{2}z -\sin\frac{\theta_2}{2}}{\sin\frac{\theta_2}{2} z+ \cos \frac{\theta_2}{2}}\Big)$ & $\{\pm i\}$ &  Rotation about $Y$-axis \& similar as $Y$-gate\\
$\begin{bmatrix}
		  e^{\frac{-i\theta}{2}}   & 0 \\ 
				0     						 & e^{\frac{i\theta}{2}} 
		\end{bmatrix}$ & \begin{quantikz} &\gate{R_Z(\theta)}& \end{quantikz} & $\psi(e^{i\theta_3} z)$ & $\{\infty,0\}$ &  Rotation about $Z$-axis \& similar as $Z$-gate\\
\bottomrule
\end{tabularx}
\end{table}




\newpage
\section{Conclusion and Further Outlook}
In this work, we have introduced a wave mechanics formalism for qubit geometry based on holomorphic functions and Mo\"bius transformations. By treating the Riemann sphere $\mathbb{C}\text{P}^1$ as a non-cotangent bundle phase space, we applied holomorphic quantization to construct a natural representation of quantum states. This approach led to a formulation of spin angular momentum operators that reproduce the standard $SU(2)$ algebra while providing new geometric insights into the evolution of quantum state. We demonstrated how fundamental single-qubit gates\footnote{Such as the Hadamard ($H$), Pauli $(X,Y,Z)$, and identity, $I$} act as Mo\"bius transformations on holomorphic wavefunctions. This interpretation offers a novel geometric perspective on quantum computation, shedding light on the implications of geometric properties of M\"obius transformations on quantum gates and their corresponding eigenstates. The results can be translated to just how quantum theory benefits from multiple equivalent formulations, that is operator, path integral, or phase-space, and quantum information can similarly benefit from holomorphic and geometric reformulations that reveal structural insights invisible in standard matrix representations.\newline

For future outlook, firstly, on the quantum gates' wavefunctions, one could also raise other geometric properties of these transformations, e.g. exact 3-transitivity and invariance of lines and circles on the Riemann sphere and find out what implications they have on the logic gates. Secondly, in a more general structure, one could generalize the technique for qubit to qudit by using the generalized Hopf fibration, $$S^{1} \longrightarrow S^{2n+1} \longrightarrow \mathbb{C}\text{P}^{n},$$ via parametrization of $S^n$ to find separable coordinates for $\mathbb{C}\text{P}^n$. Here, we have lost the spherical character of its base space in general to
take advantage, hence it necessary to utilize the higher dimensional sphere $S^{2n+1}$ to find its separable coordinates for characterizing the separable coordinates on $\mathbb{C}\text{P}^{n}$ that comes from Cartan subalgebra of $SU(n+1)$ and the rest from Casimir operators of different $SU(2)$ subalgebras \cite{Boyer}. Thirdly, one knows that qubits can also combine to form higher-dimensional qudits. Mathematically,
this translates into the problem of how the multiple $\mathbb{C}\text{P}^{1}$ are combined to form the higher-dimensional complex projective spaces. Formally it is given by $$SP^n(\mathbb{C}\text{P}^{1})=\mathbb{ C}\text{P}^{1}/S_n$$ 
where $SP^n$ stands for symmetric product of $\mathbb{ C}\text{P}^{1}$ and $S_n$ is the symmetric group \cite{Mostovoy,Blagojevic}.  As such, it is interesting to further explore that the symplectic form $\sum_{j=0}^{n}dz_j\wedge d\overline{z}_j$ of $\mathbb{ C}\text{P}^n$ is invariant under any change of $z_j$ and $\overline{z}_j$ that should be reflected in the results of its quantization where physically could be used to describe quantum entanglement. 



\section*{Acknowledgements}
This work is supported by Geran Putra, Universiti Putra
Malaysia Grant (GP/2023/ 97529000).


\newpage
\begin{thebibliography}{xxx}
	\bibitem{Shor} P.W. Shor in \textit{Proceedings 35th Annual Symposium on Foundations of Computer Science (Santa Fe, NM, USA, 1994)} pp. 124- 134

    \bibitem{Harrow} A. W. Harrow  and A. Montanaro, Quantum Computational Supremacy, Nature \textbf{549}, 14 (1997)

    \bibitem{Wiesner} S. Wiesner, Conjugate coding , ACM Sigact News \textbf{15} 1 (1983)

    \bibitem{Bennett} C. H. Bennett, G. Brassard, C. Cr\'epeau, R. Jozsa, A. Peres, and W. K. Wootters, Teleporting an Unknown Quantum State via Dual Classical and Einstein-Podolsky-Rosen Channels, Phy. Rev Lett \textbf{70}, 1895 (1993)

    \bibitem{Lloyd} V. Giovannetti, S. Llyod and L. Maccone, Quantum-enhanced Measurements: Beating the Standard Quantum Limit, Science \textbf{306}, 5700 (2004)
    
	\bibitem{Chuang} M. A Nielsen, and I.L. Chuang, \textit{Quantum Computation and Quantum Information: 10th Anniversary Edition (Cambridge University Press, 2010)}.
	
	\bibitem{Kibble} T. W. B. Kibble, Geometrization of Quantum Mechanics, Comm. Math. Phys., \textbf{65}, 189–201 (1979)
	
	\bibitem{Brody} D. C. Brody, L. P. Hughston, Geometric quantum mechanics, Jour. of Geo. and Phys., \textbf{38} 1 (2002) 
	
	\bibitem{Aharanov} J. Anandan, and Y. Aharonov,  Geometry of Quantum Evolution. Phys. Rev. Lett, \textbf{65} 1697 (1990)
	
	
	\bibitem{Schilling}  A. Ashtekar, and T. A Schilling,  Geometry of Quantum Mechanics. AIP Conference Proceedings, \textbf{342},1 (1995)
	
	\bibitem{Fioresi} E. Ercolessi, R. Fioresi, and T. Weber, The Geometry of Quantum Computing, Int. J. Geom. Meth. Mod. Phys. \textbf{21}, 10 (2024) 
	
    	\bibitem{Mosseri1}  R. Mosseri, and  R. Dandoloff,  Geometry of Entangled States, Bloch Sphere and Hopf Fibrations. J. Phys. A: Math. Gen. \textbf{34},10243 (2001).
	
	\bibitem{Mosseri2} R.Mosseri, in \textit{Topology in Condensed Matter Physics (Springer Berlin Heidelberg New York, 2006)}, pages 187–203
	
	
	
	\bibitem{Bengtsson} I.Bengtsson and K. Zyczkowski, in \textit{Geometry of Quantum States - An Introduction to Quantum Entanglement 2nd Ed. (Cambridge University Press, 2017)}.
	
	
	\bibitem{Bengsston2} I. Bengtsson, J. Br\"annlund, and K.Zyczkowsi  $\mathbb{C}\text{P}^n$ or Entanglement Illustrated. Int. Jour. of Mod. Phys. A, \textbf{17},31 (2002)
	
	\bibitem{Hiley} F. Frescura and B. Hiley, Geometric Interpretation of The Pauli Spinor Am. J. Phys. \textbf{49}, 152 (1981) 
	
	\bibitem{Kus}  M. Kus, and K. Zyczkowski,  Geometry of Entangled States. Phys. Rev. A, \textbf{63}, 032307 (2001).  
	
	\bibitem{Levay} P. Levay,  The Geometry of Entanglement: Metrics, Connections and the Geometric Phase. J. Phys. A: Math. Gen. \textbf{37} 1821 (2004). 
	
	\bibitem{Isidro} J. M.  Isidro, Quantum States from Tangent Vectors. Mod. Phys Lett. A, \textbf{19}, 31 (2004).

    \bibitem{Witten} E. Witten, Geometric Quantization of Chern-Simons Gauge theory, J. Diff. Geom. \textbf{33}, 3 (1991).
    
    \bibitem{Renate} R. Loll, In \textit{Mathematical Aspects of Classical Field Theory (AMS, 1992)}, pp. 503 - 530
 
   \bibitem{Chabaud} U. Chabaud and S. Mehraban, Holomorphic Rrepresentation of Quantum Computations, Quantum \textbf{6} 831  (2022).
	
	\bibitem{Bjelakov} I. Bjelakovic and W. Stulpe, The Projective Hilbert Space as a Classic Phase Space for Nonrelativistic Quantum Dynamics, Int. J. Theor. Phys. \textbf{44} (2005) 
	
	\bibitem{Urbantke} H. Urbantke, Two-level Quantum Systems: States, Phases, and Holonomy. Am. J. Phys.,\textbf{59}, 6 (1991) 
	
	\bibitem{Lee} J. W. Lee, C.H. Kim, E.K.  Lee, J. Kim and S. Lee, Qubit Geometry and Conformal Mapping. Quantum Information Processing \textbf{1}, (2002)
	
	
	\bibitem{Isham1} C.J.Isham in \textit{Relativity, Groups and Topology II,  B.S. DeWitt and R. Stora (eds.), (North-Holland, 1984)} pp. 1061-1290.
	
	\bibitem{Woodhouse} N.M.J. Woodhouse, Geometric Quantization, 2nd Ed. (Clarendon Press, 1997).

    
    \bibitem{PhDAHAS} H.A.S Ahmad, Canonical Group Quantization on Non-Cotangent Bundle Phase Space and its Application in Quantum Information Theory, Unpublished PhD Thesis, Universiti Putra Malaysia  (2024)
    
    \bibitem{Gurarie} D. Gurarie in \textit{Symmetries and Laplacians - Introduction to Harmonic Analysis, Group Representations and Applications, (Dover, 2008)}
    
	\bibitem{Ali} S. T. Ali and M. Englis, Quantization Methods: A Guide for Physicists and Analysts, Rev.Math. Phys. \textbf{17}, 04 (2005) 
	
	\bibitem{Isham2} C. J. Isham and A. C. Kakas, A Group Theoretical Approach to the Canonical Quantisation of Gravity. I. Construction of the Canonical Group. Class. Quantum Grav., \textbf{1} 6, (1984)
	
	\bibitem{Isham3} C. J. Isham and  A. C. Kakas A Group Theoretical Approach to the Canonical Quantisation of Gravity. II. Unitary Representations of the Canonical Group.Class. Quantum Grav., \textbf{1}, 6 (1984).
	
	\bibitem{Isham4} C. J. Isham  and N. Linden  A Group Theoretic Quantisation of Strings on Tori. Class. Quantum Grav., \textbf{5} 6 (1988)
	
	\bibitem{Zainuddin} H. Zainuddin Group-theoretic Quantization of a Particle On a Torus in a Constant Magnetic Field. Phy. Rev. D, \textbf{40}, 636 (1989) 
	\bibitem{Bouketir} A. Bouketir, Group-Theoretic Quantisation on Spheres and Quantum Hall Effect. Unpublished PhD thesis, Universiti Putra Malaysia, Malaysia (2000)
	
	\bibitem{Umar} M.F Umar,  N. M. Shah and H. Zainuddin,  Two-dimensional Plane, Modified Symplectic Structure and Quantization. Jurnal Fizik Malaysia, \textbf{39}, 2 (2018).

    \bibitem{Reyes} C. Benevides, and A. Reyes in \textit{Geometric and Topological Methods for Quantum Field Theory (Cambridge University Press, 2004)}, pp. 344–367
    
	\bibitem{Silva} R. A. Silva and T. Jacobson Particle on the Sphere: Group-Theoretic Quantization in the Presence of a Magnetic Monopole. J. Phys. A: Math. Theor. , \textbf{54}, 235303 (2021)
	
	\bibitem{Silva2} R. A. Silva and T. Jacobson, Causal Diamonds in (2+1)-dimensional Quantum Gravity. Phys.Rev.D, \textbf{107}, 2 (2023)
	
	\bibitem{Andrews} P. Andrews, The Classification of Surfaces, The American Mathematical Monthly \textbf{95},9 (1988) 

    \bibitem{Vaisman} I. Vaisman, The Bott Obstruction to the Existence of Nice Polatizations, Mn. Math \textbf{92} (1981)

	
    \bibitem{Jones} G.A. Jones and I. Singerman, in \textit{Complex Functions - An Algebraic and Geometric Viewpoint (Cambridge University Press, 1987)}.
    
	
	
	\bibitem{Boyer} C.P. Boyer,  E.G. Kalnins and P. Winternitz, Separation of Variables for the Hamilton-Jacobi Equation on Complex Projective Spaces, SIAM J. Math. Anal. \textbf{16} (1985) 
	
	\bibitem{Mostovoy} J. Mostovoy, Geometry of Truncated Symmetric Products and Real Roots of Real Polynomials, Bull. Lond. Math. Soc. \textbf{30} (1998)
	
	\bibitem{Blagojevic} P. Blagojevic, V. Grujic and R. Zivaljevic, (2004), arXiv:math/0408417v1
	
	
	
	%
	
	
	
	
	
	
	
	
	
	
	
	
	
	
	%
	
	
	
	
	
	
	
	
	
	
	
	
	
\end{thebibliography}
\end{document}